\documentclass[apl,twocolumn,showpacs,groupedaddress,preprintnumbers]{revtex4-1}

\usepackage{MnSymbol}
\usepackage[dvipdfmx]{graphicx}
\usepackage{float,graphicx}% Include figure files
\usepackage{dcolumn}% Align table columns on decimal point
\usepackage{bm}% bold math
\usepackage[mathlines]{lineno}% Enable numbering of text and display math
\usepackage{ulem}
\usepackage{color}
\usepackage[version=3]{mhchem}
\usepackage{usebib}

\begin{document}

%\preprint{Resubmitted to APL Photonics}

\title[Short Title]{Ultrafast opto-magnetic effects induced by nitrogen-vacancy centers in diamond crystals}
% Force line breaks with \\

\author{Ryosuke Sakurai}
\affiliation{Department of Applied Physics, Faculty of Pure and Applied Sciences, University of Tsukuba, 1-1-1 Tennodai, Tsukuba 305-8573, Japan.}

\author{Yuta Kainuma}
\affiliation{School of Materials Science, Japan Advanced Institute of Science and Technology, Nomi, Ishikawa 923-1292, Japan.}

\author{Toshu An}
\affiliation{School of Materials Science, Japan Advanced Institute of Science and Technology, Nomi, Ishikawa 923-1292, Japan.}

\author{Hidemi Shigekawa}
\affiliation{Department of Applied Physics, Faculty of Pure and Applied Sciences, University of Tsukuba, 1-1-1 Tennodai, Tsukuba 305-8573, Japan.}

\author{Muneaki Hase}
\email{mhase@bk.tsukuba.ac.jp}
\affiliation{Department of Applied Physics, Faculty of Pure and Applied Sciences, University of Tsukuba, 1-1-1 Tennodai, Tsukuba 305-8573, Japan.}

\date{\today}% It is always \today, today,

%Author to whom correspondence should be addressed:

\begin{abstract}
The current generation of quantum sensing technologies using color centers in diamond crystals is primarily based on the principle that the resonant microwave frequency of the luminescence between quantum levels of the nitrogen-vacancy (NV) center varies with temperature, electric and magnetic fields. This principle enables us to measure, for instance, magnetic and electric fields, as well as local temperature with nanometer resolution in conjunction with a scanning probe microscope (SPM). However, the time resolution of conventional quantum sensing technologies has been limited to microseconds due to the limited luminescence lifetime. Here, we investigate ultrafast opto-magnetic effects in diamond crystals containing nitrogen-vacancy NV centers to improve the time resolution of quantum sensing to sub-picosecond time scales. The spin ensemble from diamond NV centers induces an inverse Cotton-Mouton effect (ICME) in the form of a sub-picosecond optical response in a femtosecond pump-probe measurement. The helicity and quadratic power dependence of the ICME can be interpreted as a second-order opto-magnetic effect in which ensembles of NV electron spins act as a source for the ICME. The results provide fundamental guidelines for enabling high-resolution spatial-time quantum sensing technologies when combined with SPM techniques. 
\end{abstract}

\maketitle

\section*{I. introduction}
 Nitrogen-vacancy (NV) centers, point defects in diamond, are one of the most promising physical systems for applications in quantum computing,\cite{Maurer2012} quantum information,\cite{Mizuochi2012} and quantum sensing \cite{Pelliccione2016,Sasaki2016,Clevenson2015}; these applications rely on the coherent manipulation of spin in their negatively charged (NV$^{-}$) state. In quantum sensing, for example, the negatively charged NV$^{-}$ spin state functions as a quantum magnetometer with an all-optical readout system that can operate even at room temperature.\cite{Maze2008,Balasubramanian} The use of NV$^{-}$ centers offers strong advantages over highly sensitive quantum magnetometers such as superconducting quantum interference devices (SQUIDs), which require electrical circuits and cryogenic temperatures.\cite{Jaklevic1964}
Furthermore, combining the use of these states with scanning probe microscope (SPM) techniques, quantum sensing with a spatial resolution of tens of nanometers is possible; this technology has been used not only on magnetic materials but also on biomaterials.\cite{Wang2019} Conventional quantum sensing technologies have been developed specifically for sensitivity and spatial resolution,\cite{Magdalena2020}, however the time resolution of these techniques remains in the microseconds range.\cite{Cooper2014,Tzeng2015} 
It is thus preferable to develop new quantum sensing technologies that will enable measurements with high time resolution to precisely measure magnetic and electric fields (currents), temperatures, and other phenomena that evolve on the nanometer scale and at sub-picosecond ultrafast times.\cite{Zheng2017}

Over the last two decades, diamond nonlinear photonics has evolved based on third-order optical nonlinear susceptibility, $\chi^{(3)}$, a phenomenon which governs the optical Kerr effect (OKE),\cite{Motojima2019} two-photon absorption,\cite{Dadap1991,Maehrlein2017} and the Raman effect. \cite{Kaminskii2007} 
Second-order optical nonlinear effects in diamond crystals, however, have not been examined because second-order nonlinearity is absent in the inversion symmetric diamond lattice, i.e., $\chi^{(2)} = 0$. 
Recently, we have investigated nonlinear optical effects such as OKE and nonlinear absorption \cite{shen2003principles} in diamond crystals with NV centers using a femtosecond laser.\cite{Motojima2019} 
The introduction of NV centers resulted in an enhancement of OKE with a sub-picosecond response, while the observation of second harmonic generation (SHG) suggested that enhancement in the OKE signal may be due to second-order cascading, in which dense concentrations of NV centers effectively break inversion symmetry in the near surface region of diamond.\cite{Abulikemu2021} The OKE, which is proportional to the square of the electric field amplitude, and the Pockels effect, which is proportional to the electric field amplitude, can be used to sense the electric field strength.\cite{Valdmanis1982} For sub-picosecond magnetic field sensing, however a new nonlinear opto-magnetic effect based on the spin of NV centers is necessary.

In this study, we investigated ultrafast opto-magnetic effects in diamond crystals with NV centers using femtosecond pump-probe Kerr rotation measurements to extend the time resolution of magnetic sensing to sub-picosecond time scales. In addition to the inverse Faraday effect (IFE), we have observed the inverse Cotton-Mouton effect (ICME) in diamond, which has a different helicity than IFE and exhibits a power dependence that is second-order in the magnetic field. An ensemble of NV electron spins at a NV center in diamond can be interpreted as the source of the ICME.

\begin{figure}[t]
    \begin{center}
  \includegraphics[width=8.8cm]{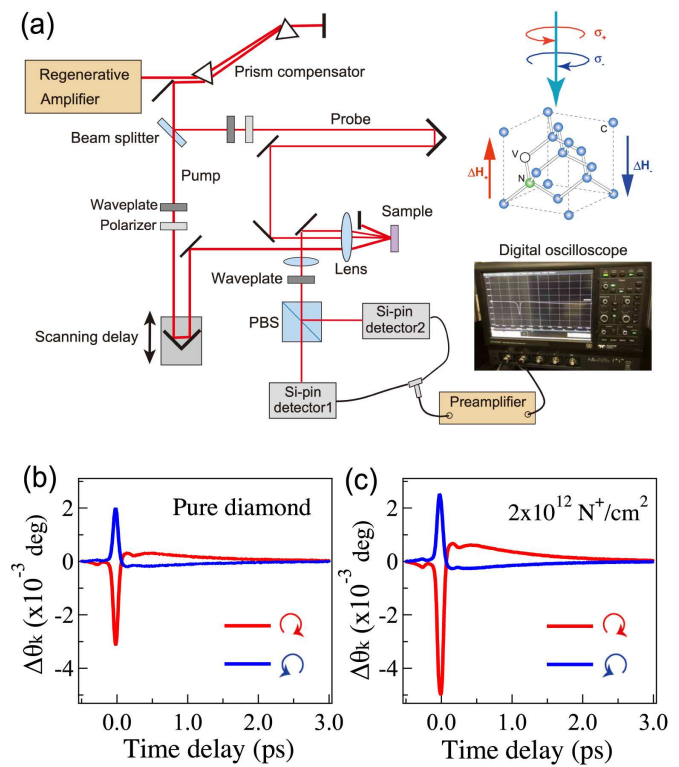}
  %  \vspace{2.5cm}
     \caption{(a) An optical setup for transient Kerr rotation measurement is illustrated. PBS represents a polarizing beam splitter cube. 
     Inset: Schematic for the Inverse Faraday effect in diamond including NV centers. The direction of the induced magnetic field $\bf{H}$ depends on the helicity of the pump light, $\pm \sigma$. (b) Time-evolution of the Kerr-rotation for Sample A at $I$ $\approx$ 29 mJ/cm$^{2}$. The symbols of $\textcolor{red}\lcirclearrowdown$ and $\textcolor{blue}\rcirclearrowdown$ denote right- and left-handed circularly polarized photons, respectively. (c) The same of (b) but for Sample C. 
     }
     \label{fig1}
    \end{center}
\end{figure}

Several studies of magneto-optical effects related to NV centers in diamond crystals have been carried out. For example, the magneto-optical Faraday effect was used to demonstrate nondestructive single-spin measurement in diamond.\cite{Buckley2010} Furthermore, the magneto-optical Voigt effect in a paramagnetic diamond membrane, with a high concentration of negatively charged NV$^{-}$ centers, has been recently demonstrated using a CW green laser, suggesting that a sub-ensemble of NV$^{-}$ centers plays a central role in the rotation of the incident linearly polarized light \cite{Haitham2020} provided that the electron spins associated with the NV$^{-}$ centers can be optically polarized into the $|0>$ ground state.\cite{Manson2006} 
As a result, while there have been numerous studies on nonlinear OKE \cite{Almeida2017,Motojima2019} and magneto-optical effects \cite{Buckley2010,Haitham2020} in pure diamond and NV-diamond, the effects of NV centers on the {\it inverse} magneto-optical effect, that is the opto-magnetic effect,\cite{Juraschek2021} have not yet been examined. An understanding of this opto-magnetic effect is, however, critical for the pursuit of further research into novel functionality of diamond nonlinear photonics so as to advance quantum sensing technology using magnetic fields or spin.
In general, the IFE can induce a transient magnetization $\bf{H}$ via irradiation by circularly polarized pump pulses, \cite{shen2003principles,kimel2005ultrafast} and can be measured by observing the rotation of a linearly polarized probe pulse transmitted through (or reflected from) a medium,\cite{kimel2005ultrafast} 
\begin{equation}
\bf{H}_{IFE} = i \alpha(\overrightarrow{E} \times \overrightarrow{E}^{*}), 
  \label{eq1}
\end{equation}
where $\alpha$ is the first-order magneto-optical coefficient and $\overrightarrow{E}$ is the electric field. 
The IFE can be observed in any material, where the third-order nonlinear optical susceptibility describing the two-photon mixing process plays a central role, similar to impulsive stimulated Raman scattering,\cite{Kirilyuk2010} where one photon is absorbed and scattered via an intermediate state, before arriving at the final spin state. 
On the other hand, the transient magnetization $\bf{H}$ induced via ICME can be expressed by, \cite{Juraschek2021}
\begin{equation}
\bf{H_{ICME}} = i\beta(\overrightarrow{E} \cdot \overrightarrow{E}^{*}) \overrightarrow{M}, 
  \label{eq2}
\end{equation}
where $\beta$ is the second-order magneto-optical coefficient and $\overrightarrow{M}$ is the applied magnetic field or magnetization. 
In the following, we report on ultrafast opto-magnetic effects, including IFE and ICME (or inverse Voigt effect), in diamond single crystals with NV centers using 800 nm and 40 fs light pulses [Fig. \ref{fig1}(a) inset].

\section*{II. Experimental}
We measured time-resolved Kerr-rotation using a reflection-based pump-probe technique, shown schematically in Fig. \ref{fig1}(a).
The light source used was a femtosecond regenerative amplifier system, which generated $\approx$40 fs pulses at a central wavelength of $\approx$800 nm with an average power of $\geq$500 mW at a 100 kHz repetition rate. The pump power was varied from 15 to 75 mW, while the probe power was maintained at a constant $\leq$2 mW. The pump and the probe beams were co-focused via a lense with a focal length $f$ = 100 mm to a spot size of $\approx$50 $\mu$m and incident angles of $\approx$20$^{\circ}$ and 25$^{\circ}$  with respect to the surface normal, respectively. The estimated pump fluence $I$ was varied from 8.0 to 40.0 mJ/cm$^{2}$. 
The probe beam was $p$-polarized while the polarization of the pump beam was varied from linearly polarized (0$^{\circ}$), to right-handed circularly polarized (45$^{\circ}$), to linearly polarized (90$^{\circ}$), to left-handed circularly polarized (135$^{\circ}$) and back to linearly polarized (180$^{\circ}$) by varying the quarter-wave-plate (QWP, $\lambda/4$ plate) angle.\cite{wilks2004transient,wilks2003investigation} The change in the Kerr-rotation of the probe pulse ($\Delta\theta_{k}$) was measured using balanced silicon photo-diodes as a function of the pump-probe delay for times up to 15 ps introduced by a shaker operating at 10 Hz.\cite{hase2012frequency} The measured signal was averaged over 500 scans to improve the signal to noise ratio. The measurements were carried out in air at room temperature.

The samples used were Element Six [100] type-IIa diamond single crystals fabricated by chemical vapor deposition, with impurity (nitrogen: [N], boron: [B]) levels  [N] $<$ 1 ppm and [B] $<$ 0.05 ppm (Sample A). The sample size was 3.0 mm$\times$3.0 mm$\times$0.3 mm (thickness). To introduce NV centers into diamond single crystals, 30 keV nitrogen ions 
($^{14}$N$^{+}$) were implanted into the diamond samples with the doses of 2.0$\times$10$^{11}$ ions/cm$^{2}$ (Sample B) and 1.0$\times$10$^{12}$ ions/cm$^{2}$ (Sample C). The implanted depth deduced from a Monte Carlo calculation was about 30--40 nm (Ref. \cite{kikuchi2017long}). 
Following implantation, the samples were annealed at 900$^{\circ}$C--1000 $^{\circ}$C in an argon atmosphere for 1 hour to produce NV centers with a production efficiency of $\approx$10$\%$.\cite{pezzagna2010creation} 
The presence of NV centers was observed by an optically detected magnetic resonance technique using a 532 nm laser at $\leq$ 1 mW,\cite{kikuchi2017long} in which it was confirmed that the electronic state of the NV diamond was dominated by the negatively charged state (NV$^{-}$) with a minor contribution from the neutrally charged state (NV$^{0}$).\cite{Abulikemu2021} 

\section*{III. Results and discussion}

Figure \ref{fig1}(b) shows the transient Kerr rotation signal of Sample A (Pure diamond) after being excited with a pump pulse of 29 mJ/cm$^{2}$ fluence, which was measured by scanning the pump-probe delay and recording the intensity difference between the orthogonal polarization components of the reflected probe light.\cite{hsieh2011selective}
The $\Delta\theta_{k}$ signal changes the sign when the helicity of the pump pulse is reversed from left- to right-handed circular polarization, which is denoted by $\textcolor{blue}\rcirclearrowdown$ and $\textcolor{red}\lcirclearrowdown$, respectively as shown in Fig.\ref{fig1}(b).
The photoexcitation of spin-polarized electrons, causes an instantaneous change in the $\Delta\theta_{k}$ signal with a near-zero time delay, which is known as the IFE.\cite{kimel2005ultrafast,hsieh2011selective,wang2016unraveling}
As seen in Fig. \ref{fig1}(c) the IFE signal was enhanced after the NV centers were introduced. 
  
\begin{figure}[t]
    \begin{center}
  \includegraphics[width=8.8cm]{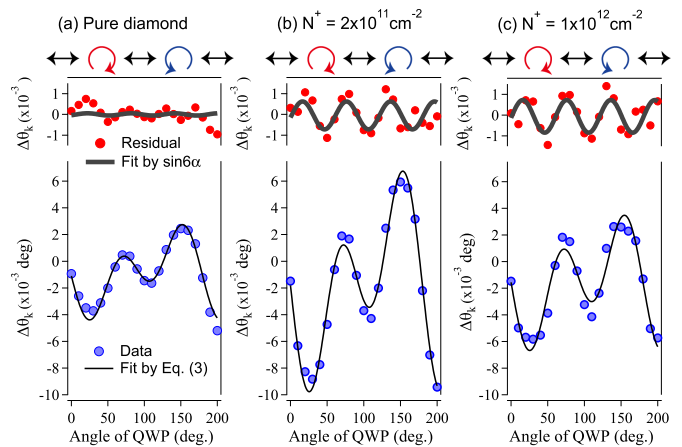}
  %  \vspace{0.3cm}
     \caption{The peak amplitude of the $\Delta\theta_{k}$ signal as a function of the pump-polarization state (helicity) for different diamond samples, (a) pure diamond (Sample A), (b) 2.0$\times$10$^{11}$ ions/cm$^{2}$ (Sample B), and (c) 1.0$\times$10$^{12}$ ions/cm$^{2}$ (Sample C). The data were taken at the pump fluence of $I$ $\approx$ 29 mJ/cm$^{2}$. 
    In the lower panel, the data are fitted using Eq. (\ref{eq3}) as shown by solid lines. The corresponding residual signals are shown in the upper panel and they are fitted using the function of $F\sin 6\alpha$ as shown by the thick solid lines.}
     \label{fig2}
    \end{center}
   \end{figure}

To investigate the effect of NV centers in more detail, we have measured the pump-polarization (helicity) dependence of the $\Delta\theta_{k}$ signal in a doped and undoped diamond samples. Figure \ref{fig2} shows the peak amplitude of the $\Delta\theta_{k}$ signal for diamond samples with and without NV centers, which exhibit a sinusoidal variation in intensity with pump polarization. The data were fit using Eq. (\ref{eq3}), which is commonly used for Kerr rotation measurements,\cite{Popov1996,wilks2004transient,wilks2003investigation} to extract the contributions from the $\sin2\alpha$ and $\sin4\alpha$ components, corresponding to the IFE and OKE signals, respectively.

\begin{equation}
\Delta\theta_k =C\sin 2\alpha + L\sin 4\alpha + D,
  \label{eq3}
\end{equation}
where $\alpha$ is the angle of the QWP, $C$ and $L$ represent the magnitude of IFE and OKE, respectively, and $D$ is a polarization-independent background. 
Interestingly, there is significant residual between the data and the fits in Fig. \ref{fig2} (b) and (c), whereas the residual is negligible in Fig. \ref{fig2}(a). 
The residuals were successfully incorporated into the fit by use of the equation $\Delta\theta'_{k} = F\sin 6\alpha$, as shown in the top panel of Fig. \ref{fig2},     
where $F$ is the magnitude and the smaller periodicity of $\sin 6\alpha$ implies the existence of a higher-order nonlinear opto-magnetic effect.\cite{Juraschek2021}

\begin{figure}[t]
   \begin{center}
 \includegraphics[width=7.0cm]{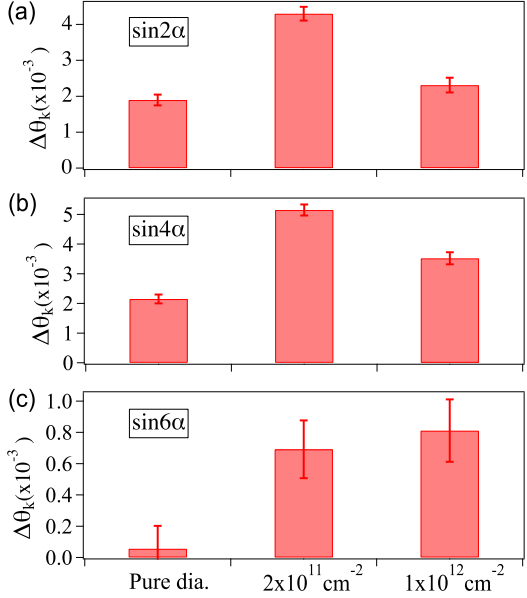}
  %  \vspace{0.3cm}
    \caption{The magnitude of (a) "$C$", (b) "$L$", and (c) "$F$" in the Kerr-rotation signal obtained for different samples, Sample A (Pure dia.), Sample B (2.0$\times$10$^{11}$ ions/cm$^{2}$), and Sample C (1.0$\times$10$^{12}$ ions/cm$^{2}$).
     }
      \label{fig3}
   \end{center}
   \end{figure}

\begin{figure}[t]
    \begin{center}
 \includegraphics[width=8.8cm]{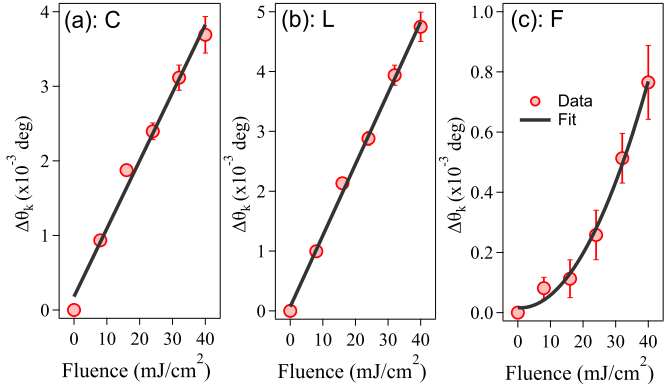}
  %  \vspace{0.3cm}
     \caption{Pump fluence dependence of the Kerr-rotation signal: The magnitude of (a) "$C$", (b) "$L$", and (c) "$F$" for sample C (1.0$\times$10$^{12}$ ions/cm$^{2}$). 
     The solid lines in (a) and (b) represent linear fits, while that in (c) represents a quadratic fit. }
     \label{fig4}
    \end{center}
   \end{figure}

To investigate the effect of NV centers on each periodic component, the magnitudes of $C$, $L$, and $F$ for each sample have been plotted in Fig. \ref{fig3}. 
As can be seen in Fig. \ref{fig3}(a) and (b), the values of $C$ and $L$ increase for Sample B, a trend that can be explained as being due to an enhancement in the cascading OKE caused by symmetry breaking around NV centers.\cite{Motojima2019,mondal2018topological,suzuki2019photon,mondal2018cascading} 
 Note that the magnitude of the Kerr rotation signal ($C$ and $L$) decreased for the highest NV concentration sample (Sample C). This observation is due to the slight loss of pump pulse energy from SHG,\cite{Abulikemu2021} and perhaps more significantly due to the modification of nonlinear optical susceptivity, $\chi^{(2)}$ and $\chi^{(3)}$, and magneto-optical susceptivity, $\chi$, as partly observed for the nonlinear refraction coefficient $n_{2}$ ($\sim$ Re$\chi^{(3)}$) in our previous pump-probe reflectivity measurements.\cite{Motojima2019}
Interestingly, the value of $F$ dramatically increases as the number of NV centers increases, i.e., from Sample A to C in Fig. \ref{fig3}(c), suggesting that the $F\sin 6\alpha$ component is related to the number of NV centers. 

To further investigate the origin of the $\sin6\alpha$ term, the $\Delta\theta_{k}$ signal intensities for each component, $C$, $L$, and $F$, are plotted as a function of the pump fluence in Fig. \ref{fig4}. 
As can be seen in Fig. \ref{fig4}(a) and \ref{fig4}(b), the linear dependence on the pump fluence indicates that the $\sin2\alpha$ and $\sin4\alpha$ periodicities as a function of helicity are consistent with the expected behavior for the IFE and OKE [see Eq. (\ref{eq1})].\cite{kimel2005ultrafast,mondal2018cascading,Motojima2019} 
In contrast, the quadratic nature of the $\sin6\alpha$ component obtained in Fig. \ref{fig4}(c) indicates an origin from a second-order opto-magnetic effect, i.e., $\Delta\theta'_{k} \propto M^{2} \propto E^{4} \propto I^{2}$ [see Eq. (\ref{eq2}) and Refs. \cite{Baranga2011,Majedi2020,Juraschek2021}]. 
The additional $\sin6\alpha$ component is found to be enhanced for the NV diamond sample for the reasons given below. 

First, as the angle of incidence to the sample is not along the surface normal, the DC magnetic field $\bf{H}$$_{IFE}$ generated via IFE has a component that is perpendicular to the beam propagation direction, as indicated by the wavevector $\bf{k}$$_{i}$. Under the condition that a DC magnetic field exists orthogonal to $\bf{k}$$_{i}$, the ICME can occur simultaneously just after the IFE. 
Using Snell's law, the refracted angle of the pump beam was calculated to be $\approx$8$^{\circ}$ from the surface normal, resulting in a perpendicular component $\bf{H}_{\perp}$$_{IFE}$ that is $\approx$14 \% of $\bf{H}$$_{IFE}$. 
Because the magnitude of the $\sin6\alpha$ component is nearly one-order of magnitude smaller than that of the $\sin2\alpha$ component (IFE), the perpendicular component $\bf{H}_{\perp}$$_{IFE}$ may play an important role in the ICME. 
Second, we propose a two-step process involving the circularly- and linearly-polarized pump light (i.e., ellipsoidally polarized light), in which the DC magnetic field $\bf{H}$$_{IFE}$ produced by circularly-polarized pump photons acts on the spin ensemble originating from diamond NV centers, 
resulting in macroscopic magnetization $\overrightarrow{M}$$_{NV}$, so that $\bf{H}$$_{ICME}$ $\propto$ $(\overrightarrow{E} \cdot \overrightarrow{E})$$\overrightarrow{M}$$_{NV}$ (Ref. \cite{Juraschek2021}). 
Here the perpendicular component $\bf{H}_{\perp}$$_{IFE}$ is expected to play a dominant role in the electronic NV spin polarization.  
The existence of a quadratic component supports our hypothesis that the $F\sin 6 \alpha$ term originates from a two-step process, in which circularly-polarized photons generate a DC magnetic field via the photon helicity of $\sin2\alpha$, and linearly-polarized photons induce transient ICME via 
$\cos4\alpha$.
To examine the possibility of the above hypothesis, we discuss the dynamics of the magnetization via NV centers in diamond. 
Figure 5 schematically shows the dynamics of the ICME (or inverse Voigt effect)\cite{Haitham2020} induced in a diamond single crystal. 
 
   \begin{figure}[t]
    \begin{center}
 \includegraphics[width=8.0cm]{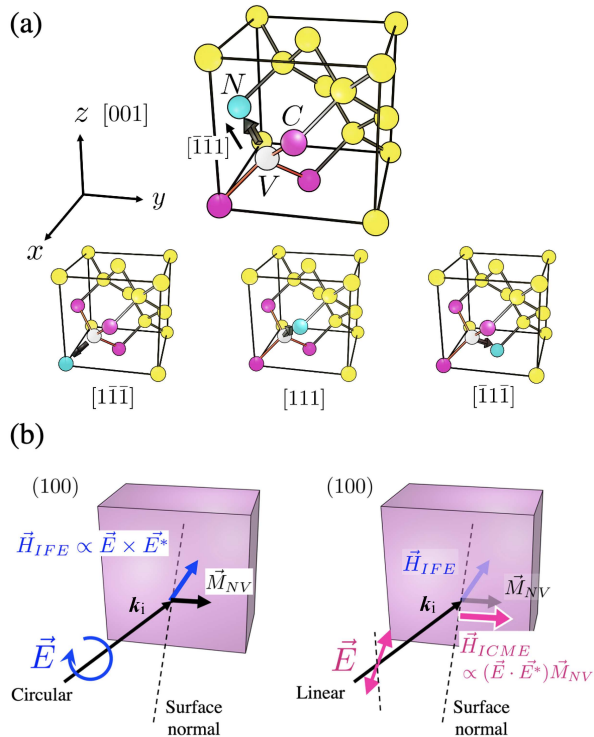}
     \caption{(a) Four possible orientations of nitrogen-vacancy (NV) color centers in diamond. Carbon atoms are depicted in yellow, nitrogen (N) atoms in blue, and vacancies (V) in white. NV electronic spin is indicated by black thick arrows. The carbon atoms which belong to the NV center are depicted in pink for the sake of clarity. (b) Schematic of the ICME (or inverse Voigt effect). On the left hand side the circularly polarized pump light generates a DC magnetic field \textbf{H$_{IFE}$} via IFE. The \textbf{H$_{IFE}$} in turn induces a coherent ensemble of NV spins. On the right hand side, the linearly polarized pump light can give rise to the ICME due to the perpendicular component of the magnetic field 
     \textbf{M$_{NV}$}, resulting from a coherent ensemble of NV spins.
     }
     \label{fig5}
    \end{center}
   \end{figure}

As shown in Fig. 5(a), due to the fcc diamond crystal structure, NV centers can orient with their symmetric axes along any of the four crystallographic axes: $[111]$, $[1\bar{1}\bar{1}]$, $[\bar{1}\bar{1}1]$, and $[\bar{1}1\bar{1}]$.\cite{Pham2012} In most diamond samples, NV centers are equally distributed across these four orientations. 
The direction of NV electronic spin is defined by the spin quantum number $m_{S}$ which has three possible values ($m_{S} = 0, \pm1$). 
As an example, we show the direction of the $m_{S} = +1$ state for each NV in Fig. \ref{fig5}(a). In thermal equilibrium, the NV  spins point in random directions, similar to those in paramagnetic materials.\cite{ashcroft2011solid}
The NV electronic spin is also oriented along these axes as indicated by black thick arrows in Fig. 5(a).\cite{Pham2012}
As shown in Fig. \ref{fig5}(b) the IFE can be induced by circularly polarized pump pulses in diamond crystals with a helicity of $\sin2\alpha$ generating a magnetic field $\textbf{H}$$_{IFE}$. 
For the case of the NV diamond crystal, the magnetic field $\textbf{H}_{IFE}$ induces a coherent ensemble of NV spins, resulting in a spin-induced macroscopic magnetic field $\overrightarrow{M}$$_{NV}$.  
This phenomenon can be explained at the macroscopic level as an impulsive stimulated Raman scattering process, involving transitions between the $m_{S} = -1$ and $m_{S}= +1$ states via an intermediate state.\cite{reid2010investigation,jin2011femtosecond}
The perpendicular component of this spin-induced magnetic field $\overrightarrow{M}$$_{NV}$ can then induce the ICME with linearly polarized light. 
In the present case, the presence of ellipsoidal polarization (circular plus linear) can play a dominant role in the above mentioned two-step process. 
Thus, the $\sin 6 \alpha$ term can be explained by combining the IFE via $\sin2\alpha$ and ICME via $\cos4\alpha$, i.e., $\sin2\alpha \cos4\alpha \sim \sin6\alpha $.
Note that while it might at first appear possible to observe the $\sin2\alpha$ term originating from the ICME, the $\sin2\alpha$ term overlaps with the periodicity of the IFE signal. Since the magnitude of the $\sin2\alpha$ term due to the ICME is an order of magnitude smaller than that of the IFE, the effect of the ICME term on the IFE signal is expected to be undetectable in the present study.

Because we did not use an external magnetic field in this study, the electronic spins of the NV centers are oriented in different directions, i.e., the four possible crystallographic axes, before light irradiation. If we place the NV diamond crystal directly above the stray magnetic field of the sample,\cite{Pelliccione2016} the $m_{S} = \pm1$ levels will split due to the Zeeman effect, and the occupation numbers of the $m_{S} = -1$ and $m_{S}= +1$ states will change depending on the spin axis direction relative to the external magnetic field.\cite{Drake2015}
As a result, NV spins in a specific direction produce a transverse component of macroscopic magnetization before light irradiation. Such a "pre-" orientation would change the magnitude of the transverse magnetic field required for the ICME, resulting in modulation of the $\Delta\theta_{k}$ intensity of the $\sin 6 \alpha$ component. By observing this modulation with a pump-probe method,\cite{Yoshida2014} 
 it may be possible to measure, e.g., the dynamics of the magnetic domain wall,\cite{Stein2013} the reversal of magnetization,\cite{Bossini2016} and current pulses \cite{Elezzabi1996} with sub-picosecond time resolution,\cite{Zhang2020} 
under realistic device-operation conditions, such as in electric circuits, power devices, topological circuits, and nano-biomaterials. 
Our approach is all-optical and is different from other conventional magnetic sensors such as SQUIDs, in which longer coherence time enables the detection of a magnetic field with high sensitivity but with a long time constant with a bandwidth in the $\leq$ MHz range.\cite{Huber2001} 
The Kerr rotation by the ICME is given by, $\Delta\theta'_{k} \approx (\chi'/n)|\overrightarrow{M}|^{2}$ (Refs. \cite{shen2003principles,kimel2005ultrafast}). Here $\chi'$ is the second-order magneto-optical susceptibility, which is unknown parameter but in general smaller than the first-order $\chi$, $n$ is the index of refraction. The magnitude of magnetization $|\overrightarrow{M}|$ by IFE in diamond can be estimated from $\Delta\theta_{k} \approx$ 1.9$\times$10$^{-3}$ degrees for the $\sin2\alpha$ term in pure diamond [Fig. \ref{fig3}(a)] to be $\sim$ 1.0$\times$10$^{4}$ Oe (or 1 T). We obtain the value $\chi'  \approx$ 3.2$\times$10$^{-3}$ deg/T$^{2}$ from the observed $\Delta\theta'_{k} \approx$ 8.0$\times$10$^{-4}$ degrees for the $\sin6\alpha$ term in NV diamond [Sample C in Fig. \ref{fig3}(c)]. Using this, and the sensitivity of our pump-probe setup ($\Delta\theta_{k} \leq$ 1.0$\times$10$^{-6}$ degrees), we estimate the possible detectable magnetic field strength to be $|\overrightarrow{M}| \approx$ 3.5$\times$10$^{2}$ Oe (or 35 mT). 
Furthermore, the signal strength of the $\sin 6 \alpha$ component may be enhanced by the selective alignment of the NV centers.\cite{fukui2014perfect}
In addition, when combined with conventional SPM techniques, the above technique may pave the way for new quantum sensing with high spatial resolution.\cite{degen2008scanning,chernobrod2005spin,hall2009sensing}

\section*{IV. conclusion}
In conclusion, we have explored the dynamics of ultrafast ICME induced by a spin ensemble from diamond NV centers in addition to the IFE and OKE. 
We discovered that the polarization dependence of the ICME, IFE, and OKE can be separated by taking into account the helicity dependence. It was also found that the $\sin 6 \alpha$ term originates from the combination of the IFE through the $\sin2\alpha$ term and the ICME through the $\cos4\alpha$ term, i.e., $\sin2\alpha \cos4\alpha \sim \sin6\alpha$. The quadratic nature of the $\sin6\alpha$ component supports the idea that this additional term is a consquence of a second-order opto-magnetic effect,  i.e., the ICME. 
As a result, we have been able to realize and experimentally demonstrate sub-picosecond opto-magnetic effects in NV center  diamond crystals. It is expected that by applying the ICME mode to nonlinear opto-magnetic quantum sensing, it will be possible to measure the local magnetic field as well as the spin current in advanced materials with high spatial-temporal resolution enabling the observation of dynamic changes in the magnetic field or spin under realistic device-operating conditions. 

\section*{Authors Contributions}
R.S. performed the experiment and analyzed the data. Y.K. and T.A. prepared the diamond samples. All authors contributed to the discussion. R.S. and M.H. prepared the manuscript.

\begin{acknowledgments}
This work was supported by CREST, JST (Grant Number. JPMJCR1875), and JSPS KAKENHI (Grant Number. 17H06088), Japan. We acknowledge Paul Fons for critically reading manuscript and stimulating discussions. 
\end{acknowledgments}

\section*{Data availability}%
The data that support the findings of this study are available from the corresponding author upon reasonable request.

%\nocite{*}

\bibliography{sakuraireference}% Produces the bibliography via BibTeX.

\pagebreak
%\newpage
%FIGURE CAPTIONS

\end{document}